\documentclass[]{sigchi-ext}

\usepackage[T1]{fontenc}
\usepackage{textcomp}
\usepackage[scaled=.92]{helvet} %
\usepackage{graphicx} %
\usepackage{balance}  %
\usepackage{ccicons}  %
\usepackage{ragged2e} %

\usepackage{subcaption}
\usepackage{todonotes}
\usepackage{amsmath}

\DeclareMathOperator*{\argmin}{arg\,min}

\def\plaintitle{Inferring Human Observer Spectral Sensitivities from Video Game Data}

\def\emptyauthor{}
\def\plainkeywords{Human Vision; Metamerism; User Study;}

\title{\plaintitle}

\numberofauthors{3}

\author{%
    \alignauthor{%
        \textbf{Chatura Samarakoon}\\
        \affaddr{Department of Engineering} \\
        \affaddr{University of Cambridge} \\
        \affaddr{Cambridge, UK.  CB3 0FA} \\
        \email{cts32@cam.ac.uk}
    }\vfil
    \alignauthor{%
        \textbf{Gehan Amaratunga}\\
        \affaddr{Department of Engineering} \\
        \affaddr{University of Cambridge} \\
        \affaddr{Cambridge, UK.  CB3 0FA} \\
        \email{gaja1@cam.ac.uk}
    }\vfil
    \alignauthor{%
        \textbf{Phillip Stanley-Marbell}\\
        \affaddr{Department of Engineering} \\
        \affaddr{University of Cambridge} \\
        \affaddr{Cambridge, UK.  CB3 0FA} \\
        \email{ps751@eng.cam.ac.uk}
    }\vfil
}

    \definecolor{linkColor}{RGB}{6,125,233}
    \hypersetup{%
      pdftitle={\plaintitle},
      pdfauthor={\emptyauthor},
      pdfkeywords={\plainkeywords},
      bookmarksnumbered,
      pdfstartview={FitH},
      colorlinks,
      citecolor=black,
      filecolor=black,
      linkcolor=black,
      urlcolor=linkColor,
      breaklinks=true,
    }

\begin{document}

\maketitle
\RaggedRight{}

\vspace{-10pt}
\begin{abstract}

    With the use of primaries which have increasingly narrow bandwidths in modern displays, observer metameric breakdown is becoming a significant factor. This can lead to discrepancies in the perceived color between different observers. If the spectral sensitivity of a user's eyes could be easily measured, next generation displays would be able to adjust the display content to ensure that the colors are perceived as intended by a given observer.

    We present a mathematical framework for calculating spectral sensitivities of a given human observer using a color matching experiment that could be done on a mobile phone display.
    This forgoes the need for expensive in-person experiments and allows system designers to easily calibrate displays to match the user's vision, \textit{in-the-wild}.
    We show how to use sRGB pixel values along with a simple display model to calculate plausible color matching functions (CMFs) for the users of a given display device (e.g., a mobile phone). We evaluate the effect of different regularization functions on the shape of the calculated CMFs and the results show that a \textit{sum of squares} regularizer is able to predict smooth and qualitatively realistic CMFs.

\end{abstract}

\pagebreak
\section{Introduction}

\begin{marginfigure}
  \begin{minipage}{\marginparwidth}
    \centering
    \includegraphics[width=0.7\linewidth]{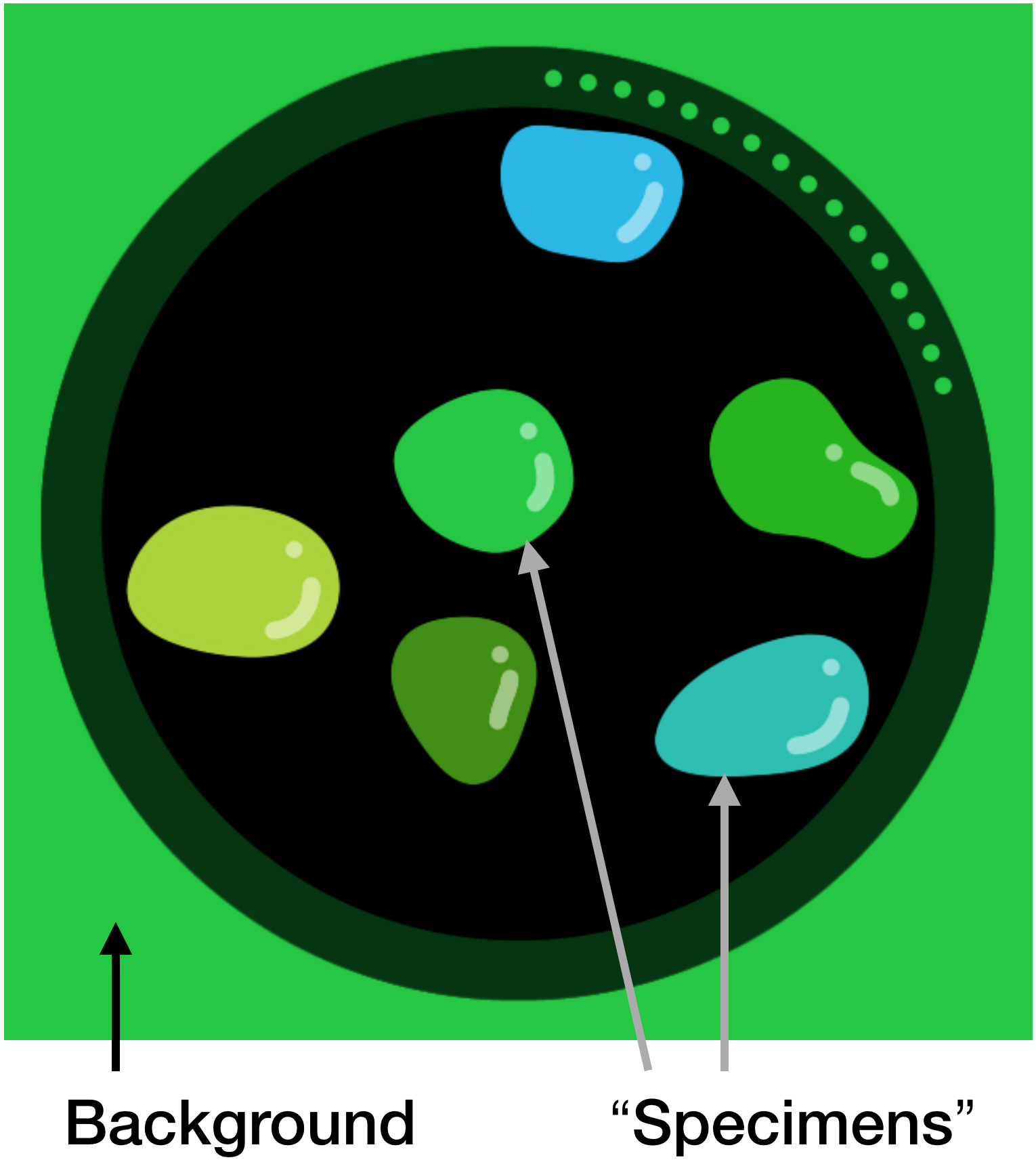}
    \caption{In Specimen, the players pick which colored blob (\textit{`specimen'}) matches the background color.}
    \label{fig:specimenScreenshot}
  \end{minipage}
\end{marginfigure}
\begin{marginfigure}
  \begin{minipage}{\marginparwidth}
    \centering
    \includegraphics[width=\linewidth]{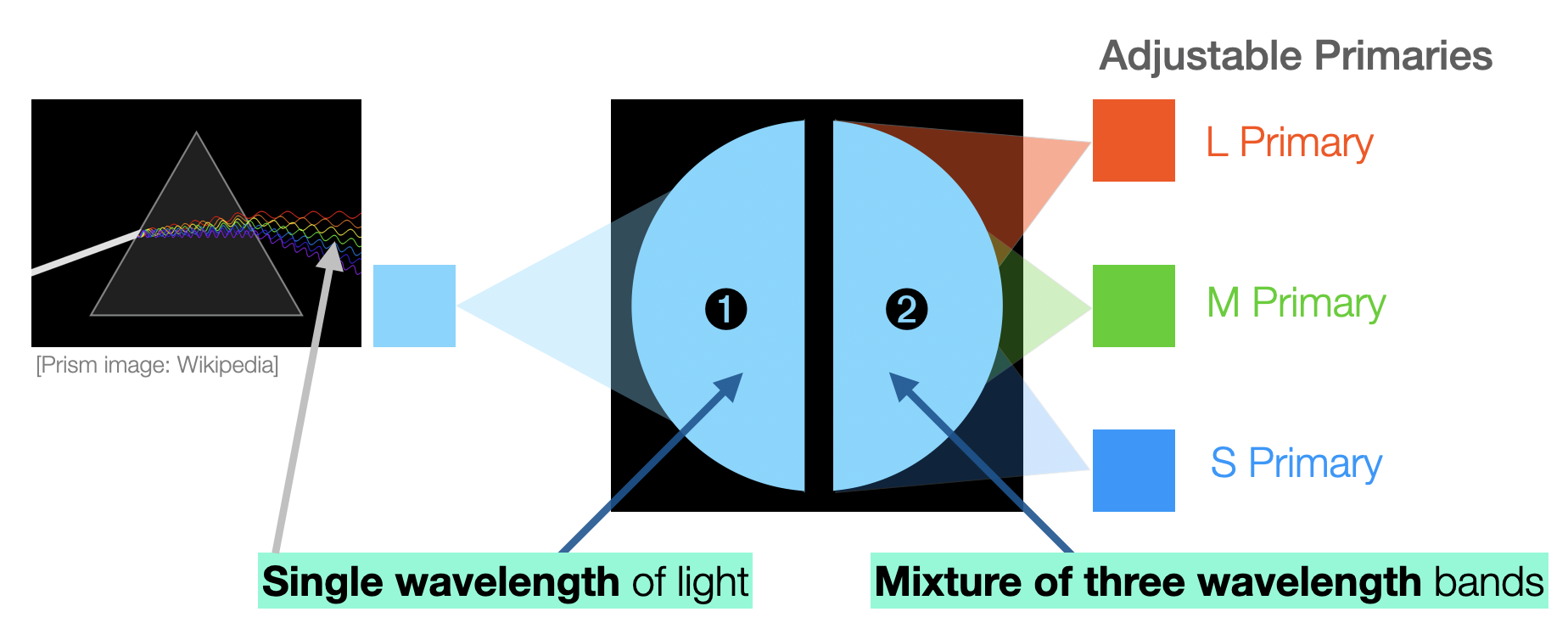}
    \caption{Traditional color matching experiments involve changing the intensities of long, medium, and short wavelength primaries till the mixture (right) is perceptually identical to the monochromatic light (left).}
    \label{fig:colorMatchExp}
  \end{minipage}
\end{marginfigure}
\begin{marginfigure}
  \begin{minipage}{\marginparwidth}
    \centering
    \includegraphics[width=\linewidth]{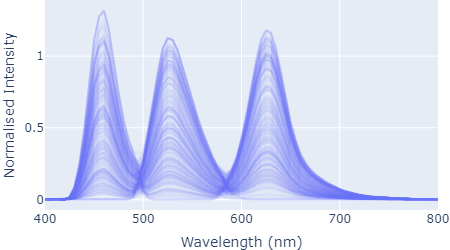}
    \caption{Due to the narrow emission bandwidths there is limited information at some wavelengths. (e.g., around 500nm and 580nm).}
    \label{fig:allSpect}
  \end{minipage}
\end{marginfigure}

Color matching experiments have been the foundation for modern colorimetry since the pioneering work by Maxwell, Young and Helmholtz~\cite{Schanda2007}. These experiments measure an individual's cone spectral sensitivities (also referred to as cone fundamentals). They involve presenting a human participant with two light sources and asking them to modulate the mixture of primaries (e.g., red, green and blue) in one light source until it visually matches a second light source comprising a single wavelength (see Figure~\ref{fig:colorMatchExp})~\cite{Guild1925,Wright1929a}. From this, the required proportions of each of the red, green, and blue primaries to match each target wavelength can be obtained. The three functions for red, green, and blue comprises the individual color matching functions (CMFs). They are usually given in the space of imaginary primaries X, Y, and Z to make the CMFs strictly positive. The CMFs can be used to numerically describe perceptual color equivalence. They can also be converted to the cone fundamentals through a linear transformation.
As might be expected, these in-person user studies are expensive and time consuming to conduct.

Modern displays are using increasingly narrower primaries. They have been shown to be more prone to observer metameric breakdown compared to displays with wideband primaries~\cite{Sarkar2010b}. Observer metameric breakdown is the process where two different observers disagree on what `color' a particular spectrum is. Thus far, standardization of color spaces and associated primaries have been used as a way to ensure color constancy across different media (e.g., displays, print media, etc.). However, \textit{observer metameric breakdown} cannot be as easily circumvented. One solution is to do a secondary display calibration to suit the sensitivities of the user's eyes. This could work for devices that primarily have a single user (e.g., mobile phones, laptops).

In this work, we explore a framework for characterizing a user's spectral sensitivities without using an in-person experiment and evaluate the results using the \textit{Specimen} dataset~\cite{Stanley-Marbell2018b}.

\subsection{The \textit{Specimen} dataset}

In 2015, \textit{PepRally} released \textit{Specimen}, a color matching game for iOS~\cite{PepRally2015a}. The game involves the players picking which color blob (`specimen') matches the background color (see Figure~\ref{fig:specimenScreenshot}). If the player choses the correct specimen they will be presented with a new background color. Incorrect choices reset the bonus streak. This repeats until all the specimens have been matched correctly. The game keeps a log of every choice a player makes through in-game analytics with details about the chosen color and the correct (background) color choice as sRGB values, and some other bits of information like device model, time, and anonymized user ID. The game sends the anonymized data to an analytics facility using Yahoo Slurry. The aggregate analytics dataset contains data from 41,000 players, totaling 489k play sessions and 28.6 million color matches.

We present the results of the preliminary investigation we carried out into using these RGB color matching data to extract the players' CMFs. We show that its possible to devise a numerical method to extract the CMFs.

\vspace{-2pt}
\subsection{Contributions}
In this work we present the following contributions.
\begin{enumerate}[noitemsep, nolistsep]
  \vspace{-1mm}
  \item A mathematical framework for extracting color vision cone response for a particular human observer (the so-called \textit{cone fundamentals}) from data obtained from a popular color matching game~\cite{PepRally2015a,STCLAIR2020} in the iOS app store.
  \item Implementation of the mathematical framework using TensorFlow’s optimizer backend.
  \item Demonstration that the implementation of the framework, applied to data from the \textit{Specimen} dataset, and across four different priors (uniform, gaussian triple, $2^\circ$ standard observer, and $10^\circ$ standard observer), provides meaningful color matching functions.
\end{enumerate}

\section{Problem Definition}
\label{sec:probDef}

Let $\overline{C}_{rgb} = [C_r,C_g,C_b]$ be an incorrectly chosen color and $\overline{T}_{rgb}= [T_r,T_g,T_b]$ be the correct target color, both in the in sRGB color space with values normalized to the range $[0,1]$.
\begin{marginfigure}
  \begin{minipage}{\marginparwidth}
    \vspace{-50pt}
    \centering
    \includegraphics[width=\linewidth,trim={0 0 0 0}, clip]{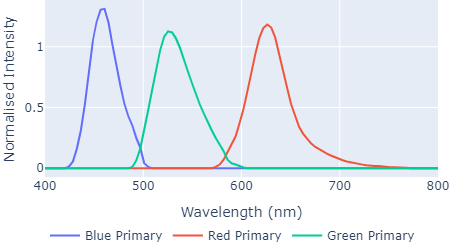}
    \caption{The shapes of the red, green and blue primary emissions of the iPhone X display were extracted from the white point spectrum.}
    \label{fig:primaries}
    \vspace{10pt}
  \end{minipage}
\end{marginfigure}
\begin{marginfigure}
  \begin{minipage}{\marginparwidth}
    \centering
    \includegraphics[width=\linewidth]{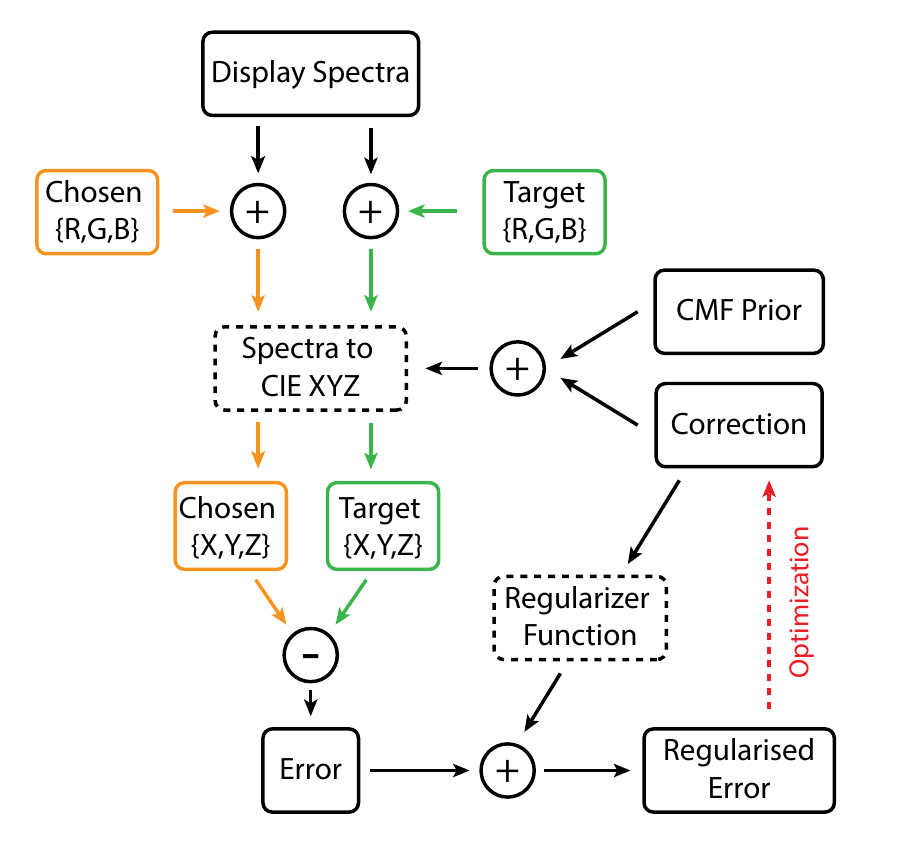}
  \end{minipage}
  \caption{Flow chart showing the architecture used to learn the spectral correction term.}
  \label{fig:calcFlowchart}
\end{marginfigure}

Also, let $\overline{P}= [P_r,P_g,P_b]$ be the spectra for the chosen display's red, green, and blue primaries.
The spectrum of the chosen color and the target color can be calculated using the Equations \ref{eqn:chosenSpect} and \ref{eqn:targetSpect} respectively.
\begin{equation}
  S_c = \sum_{k\in\{r,g,b\}} P_k \cdot C_k
  \label{eqn:chosenSpect}
\end{equation}
\begin{equation}
  S_t = \sum_{k\in\{r,g,b\}} P_k \cdot T_k
  \label{eqn:targetSpect}
\end{equation}

Let $\overline{\phi}_{std} = [\overline{x}_{std},\overline{y}_{std},\overline{z}_{std}]$ be the CMFs for $2^\circ$ or $10^\circ$ CIE standard observers. The CIE tristimulus values $[X,Y,Z]$ for the chosen spectrum $S_c$ are calculated as,
\begin{equation}
  X = \int_{\lambda} S_c \cdot  \overline{x}_{std} \ d\lambda
  \label{eqn:tristimCalc_x}
\end{equation}
\begin{equation}
  Y= \int_{\lambda} S_c \cdot  \overline{y}_{std} \ d\lambda
  \label{eqn:tristimCalc_y}
\end{equation}
\begin{equation}
  Z = \int_{\lambda} S_c  \cdot \overline{z}_{std} \ d\lambda
  \label{eqn:tristimCalc_z}
\end{equation}
Similarly for the target spectrum $S_t$.

Let $\overline{\phi}_{u}$ be the individual CMFs for the user. Also, let $\overline{\delta_u}$ be a correction term with the same domain as $\overline{\phi}_{std}$ such that $\overline{\phi}_{u} = \overline{\phi}_{std} + \overline{\delta_u}$. The correction term represents the deviation of the user's vision from the standard observer color matching functions.

The user adjusted tristimulus values can be calculated using Equations \ref{eqn:tristimCalc_x} to \ref{eqn:tristimCalc_z} with individual CMFs, $\overline{\phi}_{u}$, instead of the standard CMFs, $\overline{\phi}_{std}$. Let $\overline{\Phi^u}_c = [X_c^u,Y_c^u,Z_c^u]$  and $\overline{\Phi^u}_t =[X_t^u,Y_t^u,Z_t^u]$ be these user adjusted tristimulus values for the chosen spectrum and the target spectrum respectively.

If $\overline{\Phi^u}_c = \overline{\Phi^u}_t$ this implies that the colors are perceptually similar. The target is to find the correction term $\overline{\delta_u}$ that makes $\overline{\Phi^u}_c = \overline{\Phi^u}_t$. This can be posed as the following minimization problem,
\begin{equation}
  \argmin_{\overline{\delta_u}} \| \overline{\Phi^u}_c - \overline{\Phi^u}_t \|_2^2
\end{equation}

To prevent the correction term from causing large deviations, we explored numerous regularization functions. The regularized optimization problem is given by the following equation.
\begin{equation}
  \argmin_{\overline{\delta_u}} (\| \overline{\Phi^u}_c - \overline{\Phi^u}_t \|_2^2 + R(\overline{\delta_u}) )
  \label{eqn:conOpt}
\end{equation}
where R is a regularization function.

Equation \ref{eqn:conOpt} gives the optimization process using a single color selection event. However, the optimum value for $\overline{\delta_u}$ should be valid across all the color selection events for a given user.
Thus, for the optimization, we consider all the user's selection events in a batch and calculate the mean value for $\| \overline{\Phi^u}_c - \overline{\Phi^u}_t \| _2^2$ across this batch before evaluating the cost function given in Equation~\ref{eqn:conOpt}.
By evaluating Equation~\ref{eqn:conOpt}, we can infer the individual CMFs from the RGB values of the mismatched color pairs.

\begin{marginfigure}
  \begin{minipage}{\marginparwidth}
    \centering
    \includegraphics[width=\linewidth]{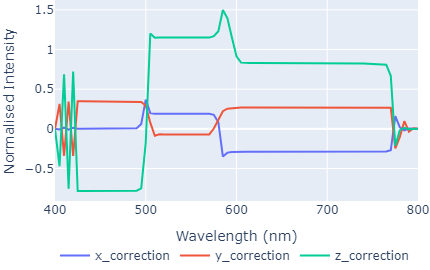}
  \end{minipage}
  \begin{minipage}{\marginparwidth}
    \centering
    \includegraphics[width=\linewidth]{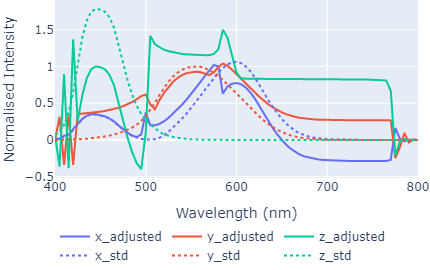}
  \end{minipage}
  \vspace{-10pt}
  \caption{Without regularization the optimization overfits and results in calculated values are physically impossible (correction (above), CMFs (below)).}
  \label{fig:noReg}
  \vspace{10pt}
\end{marginfigure}
\begin{marginfigure}
  \begin{minipage}{\marginparwidth}
    \centering
    \includegraphics[width=\linewidth]{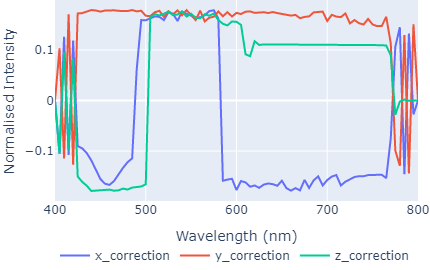}
  \end{minipage}
  \begin{minipage}{\marginparwidth}
    \centering
    \includegraphics[width=\linewidth]{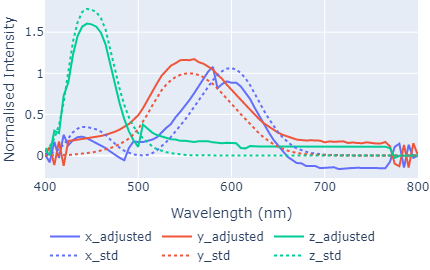}
  \end{minipage}
  \vspace{-10pt}
  \caption{\textit{Max absolute} regularization leads to slightly smoother curves than non-regularized optimization (correction (above), CMFs (below)).}
  \label{fig:MaxAbs_2d}
  \vspace{10pt}
\end{marginfigure}

\section{Methodology}

The \textit{Specimen} app stopped collecting data in 2018. At the time, the only iOS phone with an OLED display was the \textit{iPhone X}, which we used as the target platform. We use the fact that OLED displays have clearly delineated primary emission spectra in out analysis. We used the white-point spectral measurements of the display carried out by Raymond Soneira~\cite{Corporation2017} as a starting point and manually isolated the shapes of the three primary emissions from the combined white spectrum. We used this as the display model.

In the \textit{Specimen} dataset we found 141 users who used \textit{iPhone X}s totaling 21,250 color matches.
From that set of players, we chose the player with the most incorrect color matches (344 incorrect matches from 2,042 total matches) as the individual whose CMFs are to be learnt in this preliminary report.

We calculated the emission spectra for each mismatched chosen and target color pair by multiplying the normalized R, G, and B values with the corresponding subpixel emission. %

We then used Tensorflow's optimizer backend with the Adam optimizer~\cite{Kingma2015} with up to 10,000 iterations to find the CMF correction term required to make the chosen spectrum and the target spectrum, perceptually identical.

We evaluated four priors for the CMFs; namely, the $2^\circ$ and $10^\circ$ CIE standard observer CMFs~\cite{ColorandVisionResearchLab2020}, a uniform prior, and a gaussian mixture prior. For the gaussian mixture, we used the peak wavelengths and intensities from the $2^\circ$ standard observer.

Additionally, we used the following regularizers;
\begin{enumerate}[noitemsep,nolistsep]
  \vspace{-1mm}

  \item Max of absolute values : $max(|\overline{\delta_u}|)$ %
  \item Mean of Absolute values:  $\Sigma(|\overline{\delta_u}|)/n$ %
  \item Sum of Absolute values~\cite{Tibshirani1996}: $\Sigma(|\overline{\delta_u}|)$ %
  \item Root Mean Squares : $\Sigma(\|\overline{\delta_u}\|_2)/n$
  \item Mean of Squares : $\Sigma(\|\overline{\delta_u}\|_2^2)/n$  %
  \item Sum of Squares~\cite{Willoughby1979}: $\Sigma(\|\overline{\delta_u}\|_2^2)$ %
\end{enumerate}

We are using XYZ color matching functions instead of the cone fundamentals because it simplifies the optimization process. And as outlined before, cone fundamentals can be extracted from the CMFs using a known linear transformation~\cite{Stockman2019} as outlined before.

\section{Results}

\subsection{Standard Observer Priors}
\begin{figure}[h]
  \begin{subfigure}[b]{0.49\linewidth}
    \includegraphics[width=\textwidth,trim={0 0 0 0}, clip]{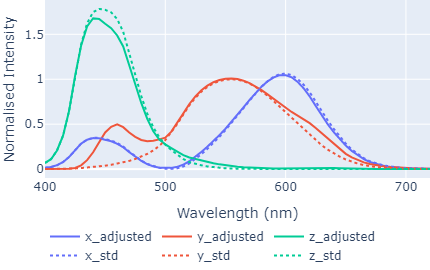}
    \caption{$2^\circ$ CMF}
    \label{fig:learntCMFs_2d}
  \end{subfigure}%
  \begin{subfigure}[b]{0.49\linewidth}
    \includegraphics[width=\textwidth,trim={0 0 0 0}, clip]{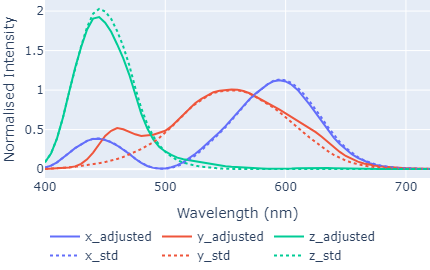}
    \caption{$10^\circ$ CMF}
    \label{fig:learntCMFs_10d}
  \end{subfigure}
  \caption{Optimizing with \textit{sum of squares} regularizer show that both $2^\circ$ and $10^\circ$ standard observer CMF has been adjusted to explain the observed color confusions better.}
\end{figure}

\begin{marginfigure}
  \begin{minipage}{\marginparwidth}
    \centering
    \includegraphics[width=\linewidth]{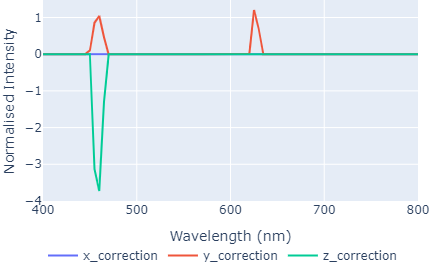}
  \end{minipage}
  \begin{minipage}{\marginparwidth}
    \centering
    \includegraphics[width=\linewidth]{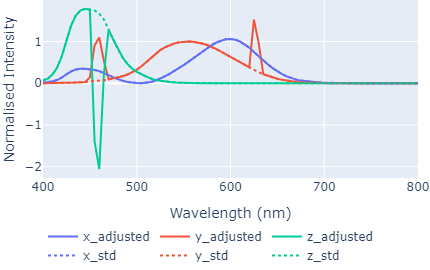}
  \end{minipage}
  \vspace{-10pt}
  \caption{\textit{Mean absolute} regularization causes \textit{peaky} artifacts (correction (above), CMFs (below)).}
  \label{fig:MeanAbs_2d}
  \vspace{10pt}
\end{marginfigure}

\begin{marginfigure}
  \begin{minipage}{\marginparwidth}
    \centering
    \includegraphics[width=\linewidth]{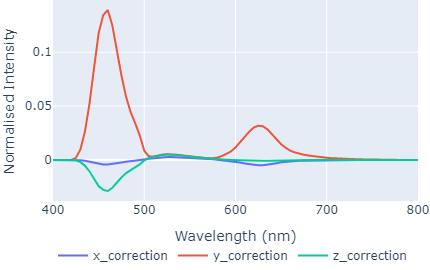}
  \end{minipage}
  \begin{minipage}{\marginparwidth}
    \centering
    \includegraphics[width=\linewidth]{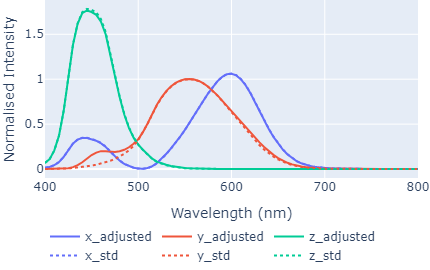}
  \end{minipage}
  \vspace{-10pt}
  \caption{\textit{Sum of Squares} regularization leads to the smoothest and most realistic looking CMFs (correction (above), CMFs (below)).}
  \label{fig:SumSquares_2d}
  \vspace{10pt}
\end{marginfigure}

\begin{marginfigure}
  \begin{minipage}{\marginparwidth}
    \centering
    \includegraphics[width=\linewidth]{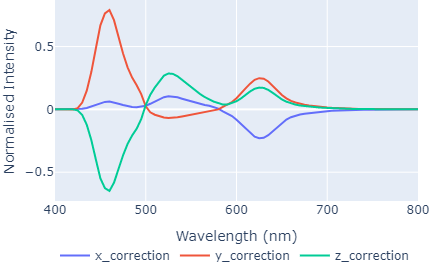}
  \end{minipage}
  \begin{minipage}{\marginparwidth}
    \centering
    \includegraphics[width=\linewidth]{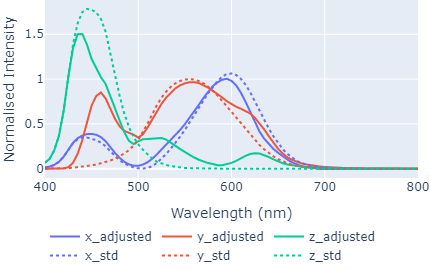}
  \end{minipage}
  \vspace{-10pt}
  \caption{\textit{Mean Squares} regularizer results in a smooth set of CMFs but show more fluctuation around the CMF prior compared to the \textit{sum of squares} regularizer (correction (above), CMFs (below)).}
  \label{fig:MeanSquares_2d}
  \vspace{10pt}
\end{marginfigure}

\vspace{-5pt}
The Figures \ref{fig:MaxAbs_2d}, \ref{fig:MeanAbs_2d}, \ref{fig:SumSquares_2d} and \ref{fig:MeanSquares_2d} on the sidebar show the qualitative results of the different regularizers for the $2^\circ$ standard observer prior. We were unable to calculate gradients when using \textit{root mean squares} regularizer and thus were unable to optimize using it. The first three regularizers, $max(|\overline{\delta_u}|)$, $\Sigma(|\overline{\delta_u}|)/n$, and $\Sigma(|\overline{\delta_u}|)$, led to CMFs that have negative values and are implausible: The CIE XYZ color space is defined such that the CMFs are strictly positive. Both $\Sigma(\|\overline{\delta_u}\|_2^2)/n$ and $\Sigma(\|\overline{\delta_u}\|_2^2)$ regularizers led to smooth CMFs with the former achieving smaller final cost.

\begin{marginfigure}
  \begin{minipage}{\marginparwidth}
    \centering
    \includegraphics[width=\linewidth,  trim={0 0 0 0}, clip]{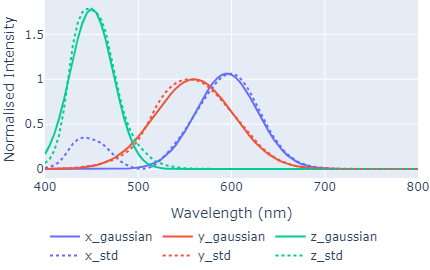}
  \end{minipage}
  \vspace{-10pt}
  \caption{$2^\circ$ standard observer CMFs were approximated with three gaussians.}
  \label{fig:gaussApproxCMFs_2d}
  \vspace{10pt}
\end{marginfigure}
\begin{marginfigure}
  \begin{minipage}{\marginparwidth}
    \centering
    \includegraphics[width=\linewidth,  trim={0 0 0 0}, clip]{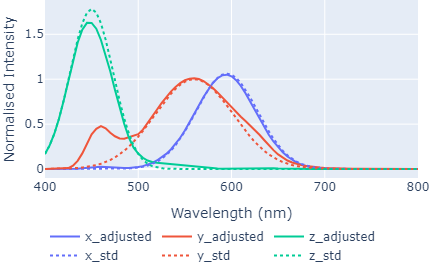}
  \end{minipage}
  \vspace{-10pt}
  \caption{The CMFs learnt with the gaussian approximation is similar to those obtained when using the $2^\circ$ standard observer prior.}
  \label{fig:gaussLearntCMFs_2d}
  \vspace{10pt}
\end{marginfigure}

\vspace{-5pt}
Overall, sum of squares ($\Sigma(\|\overline{\delta_u}\|_2^2)$) led to the smoothest and the most qualitatively realistic CMFs. Figures \ref{fig:learntCMFs_2d} and \ref{fig:learntCMFs_10d} show the results for the chosen user for both $2^\circ$ and $10^\circ$ standard observer priors using a weight of 0.05 on the regularizer term. The figures show that the training process is able to learn corrections for the standard observer CMFs that better explains the observed color confusions without resulting in unnatural CMF shapes.

\vspace{-5pt}
\subsection{Uniform Priors}
With a uniform prior, all three CMFs start off with a fixed value across the range of wavelengths.
However, the chosen color and the target color in the XYZ space is coupled together because the calculation of both those values rely on the base CMF (see Figure~\ref{fig:calcFlowchart}). This led to the optimization process resulting CMFs that are uniform, which is incorrect. We also tried adjusting the values for the CMF prior to match the peak intensities of the $2^\circ$ standard observer CMFs which produced a similarly unrealistic result. This result is not surprising but was worth evaluating to provide a complete picture.

\vspace{-5pt}
\subsection{Gaussian Priors}

With the Gaussian prior, the $2^\circ$ standard observer CMFs were approximated with 3 gaussians (see Figure~\ref{fig:gaussApproxCMFs_2d}). Figure~\ref{fig:gaussLearntCMFs_2d} shows the solution obtained with the sum of squares regularizer with a 0.05 weight (identical setup to Figures \ref{fig:learntCMFs_2d} and \ref{fig:learntCMFs_10d}). The results show a striking similarity to the results of the $2^\circ$ standard observer prior showing that our method produces consistent results.

\section{Future work}

As outlined before, this work presents a purely numerical framework for extracting CMFs. Although the results show that the optimization process results in plausible, smooth CMFs, there is no guarantee that they are physically accurate.

In addition to that, this work makes the following approximations that limit the accuracy of the results.
From the perspective of the display, we are using a simple model assuming that the spectrum of the real display emission can be obtained by using the normalized RGB values as a multiplier to rescale the peak emissions of the primaries.
We are also using normalized intensity as opposed to spectral radiance in calculating the [X,Y,Z] tristimulus values.
Furthermore, we are using the CIE XYZ space to minimize the perceived difference instead of the CIE Lab space which is better at representing perceptual differences.

As a necessary extension to this ``late breaking results'' submission, we are exploring the following avenues to forgo the approximations and make the results physically accurate.

We need three things to make our system feasible; namely, a model that incorporates physical attributes of the human eye, a more accurate display model, and a user study to validate the results.

To achieve the first goal, we aim to find the correction term in cone fundamental space instead of the XYZ CMF space. This would allow us to incorporate optical properties of the eye more easily into the regularization term. In addition to that, we aim to carry out the optimization in the CIE LAB space instead of the CIE XYZ space.

To build a better model of the display we aim to measure the variation of the emission spectrum of the display using an optical spectrometer while varying the display's RGB values. This would allow us to create an model that accurately maps from the RGB space to the spectral space.

For the user study, we hope to carry out a traditional color matching experiment to measure the users' true CMFs.  Following that we aim to have the users play a version of the \textit{Specimen} game and predict the CMFs. We can then use the measured CMFs to learn the regularization needed to ensure that the calculated CMFs match the measured CMFs. Finally, we can use this validated method to extract physically accurate CMFs from the \textit{Specimen} dataset for the 141 players.

\section{Related Work}

As outlined before, the study of human vision using color matching experiments go back to the 19th century with Maxwell,Young and Helmholtz~\cite{Schanda2007}. The modern day color matching experiments are based on Guild~\cite{Guild1925} and Wright's~\cite{Wright1929a} work in the 1920s. In their experiments, they used a set of monochromatic target sources and an controllable additive mixture of red, green and blue sources.

Conceptually our work is similar to the extent that we are also using controllable red, green blue primaries, but in the form of pixels on a opto-electronic display. But instead of having an independent target that we are trying to find a match for, both the target and the match are coupled by the properties of the display.
To our knowledge, no other work has been reported in the literature that approaches this problem numerically and at this scale. We are able to carry out this analysis because we have access to the \textit{Specimen} dataset, which is not in the public domain.

The methods we describe in this work could form the basis for new color transformations which trade display power dissipation for color fidelity~\cite{2016:CSP:2901318.2901347}. Such color approximation optimizations could in turn be combined with power-saving I/O encoding techniques~\cite{2016:RSI:2897937.2898079} or even inferring permissible color approximation from programming languages that permit programmers to specify accuracy constraints~\cite{M:pmup06}.

\section{Acknowledgements}
This research is supported by an Alan Turing Institute award TU/B/000096
under EPSRC grant EP/N510129/1. C. Samarakoon is supported by the EPSRC DTP Studentship award EP/N509620/1.

\balance{}

\footnotesize
\bibliography{sp_cmf-refs}
\bibliographystyle{SIGCHI-Reference-Format}

\end{document}